\documentclass[12pt]{iopart}

\usepackage{psfrag}
\usepackage{bm}
\usepackage{verbatim}
\usepackage{bm}
\usepackage{graphicx}  
\usepackage{latexsym}                
\usepackage{iopams}
\usepackage{amsfonts}  
\usepackage{amssymb}
\usepackage{amsthm}
\usepackage{dcolumn}
\usepackage{color}
\usepackage{mathrsfs}

\newcommand{\be}{\begin{equation}}
\newcommand{\ee}{\end{equation}}
\newcommand{\bea}{\begin{eqnarray}}
\newcommand{\eea}{\end{eqnarray}}

\def\softt{{\leavevmode\setbox1=\hbox{t}%
\hbox to \wd1{t\kern-0.6ex{\char039}\hss}}}

\begin{document}
\title{Passage-time statistics of superradiant light pulses from Bose-Einstein condensates}

\author{L.~F.~Buchmann$^{1,2}$,G.~M.~Nikolopoulos$^1$,O.~Zobay$^3$ and P.~Lambropoulos$^{1,2}$}
\address{$^1$ Institute of Electronic Structure and Laser, Foundation for Research and 
Technology-Hellas, GR-71110 Heraklion, Crete, Greece}
\address{$^2$ Department of Physics, University of Crete, P.O. Box 2208, GR-71003 Heraklion, Crete, Greece}
\address{$^3$ School of Mathematics, University of Bristol, University Walk, Bristol BS8 1TW, UK}

\ead{buchmann@iesl.forth.gr}

\begin{abstract}
We discuss the passage-time statistics of superradiant light pulses generated during the scattering 
of laser light from an 
elongated atomic
Bose-Einstein condensate. Focusing on the early-stage of the 
phenomenon, we analyze the corresponding probability distributions and 
their scaling behaviour with respect to the threshold photon number and the coupling strength. With respect to these parameters, we
find 
quantities 
which only vary significantly
during the transition between the Kapitza Dirac and the Bragg regimes. 
A possible connection of the present observations to Brownian motion is also discussed.  
\end{abstract}

\submitto{\jpb}
\maketitle

\section{Introduction}

The interaction of optical fields with Bose-Einstein condensates (BECs) has attracted 
considerable interest over the last years. A rather intriguing phenomenon in this context 
is the generation of superradiant light pulses accompanied by condensate momentum side-modes, 
when far off-resonant laser light is scattered from an elongated BEC 
\cite{InoChiSta99,VogXuKett,SRExp,SRTheorSemiclass,SRTheorQuant,MooMeyPRL,ZobNikPRA,BuNikZobLam}. 
The process is initially driven by quantum fluctuations in the vacuum electromagnetic field 
(i.e., spontaneous Rayleigh scattering) \cite{MooMeyPRL}. After the first photons have been scattered, 
bosonic stimulation sets in and additional successive scattering events give rise to 
a rapid growth of the emitted radiation and of the side-mode populations. 
Moreover, due to the geometry of the condensate, the gain is largest when the scattered 
photons leave the condensate along its long axis, in the so called endfire modes. 
The fact that the process is triggered by quantum 
fluctuations is
expected to be 
reflected in various macroscopic parameters of the superradiant (SR) light 
pulses (e.g., peak intensity, shape, etc), which should also fluctuate when the 
same experiment is repeated.

Throughout this work we focus on the passage-time statistics, which refer to the fluctuations   
in the time at which the pulse intensity at a given point in space reaches a certain reference 
intensity. The effect of quantum fluctuations is expected to be rather prominent for short 
times, where the observed intensity is relatively small. 
Given that the 
populations of the endfire modes and the side-modes grow fast in time, 
the evolution of the system becomes practically deterministic very soon \cite{BuNikZobLam}
. 
For our purposes, it therefore suffices to
focus on the early stage of the phenomenon,
where the BEC is practically undepleted, only first-order atomic side-modes are significantly 
occupied, and the dynamics of the system are governed by linearized equations of motion. 
In view of the above arguments, the passage-time statistics at the early stage are expected 
to represent adequately also the delay-time statistics, which refer to the first maximum 
of the pulse that appears at later times. 

Analogous theoretical as well as experimental studies have been performed in the context of 
conventional (Dicke) superradiance \cite{Dicke,Benedict,GroHar82}, 
where excited atoms interacting with the vacuum of the 
electromagnetic field emit collectively; analytic expressions for the probability 
distribution of the passage and delay times have been obtained \cite{GroHar82,Haake}. 
In the system under consideration, however, condensed atoms may also scatter a photon from the 
endfire modes back into the laser mode, giving thus rise to recoiling atoms 
moving against the direction of the applied laser field (backward atomic side-modes). 
This essential feature of 
superradiance from Bose-Einstein condensates
 is a key difference 
between this process and conventional superradiance. In this work we investigate how the presence of backward 
atomic side-modes affects the passage-time statistics, and examine whether the corresponding 
probability distributions usually discussed in conventional 
superradiance can also describe accurately the fluctuations in systems involving ultracold atoms. To the best 
of our knowledge, none of these questions has been addressed either theoretically or 
experimentally so far . In the following section we present the theoretical framework, 
while in Sec. \ref{secIII} we analyze the passage-time statistics and their scaling behaviour. 
We conclude in Sec. \ref{conclusion}.

\section{Theoretical Framework}
\label{secII}
The system under consideration pertains to a BEC elongated along the $z$-axis, consisting of $N$ atoms. A linearly polarized laser with frequency $\omega_l=k_l c$, far detuned from the closest atomic transition by a value of $\delta$, is illuminating the cloud along the $x$-axis. 
Due to the coherent nature of the condensate, successive Rayleigh scattering events are strongly correlated and lead to collective superradiant behaviour. 
As a result of the cigar-shape of the condensate, the gain is largest when the scattered photons leave the condensate along its long axis, traveling in the so called {\it endfire modes} with wave vectors ${\bf k}\approx \pm k_l{\bf e}_z$.
A condensate atom can scatter a laser photon into the endfire modes, experiencing a recoil $\hbar{\bf q}\approx\hbar(k_l{\bf e}_x -{\bf k})$. On the other hand, it can also  scatter a photon from the endfire modes into the laser mode, in which case its momentum changes by $\hbar{\bf q}\approx \hbar (-k_l{\bf e}_x + {\bf k})$. These processes lead to the formation of two pairs of atomic {\it side-modes}, consisting of counterpropagating atoms with a narrow momentum spread (compared to $k_l$). 
Of course, atoms within these side-modes can also scatter photons, thereby acquiring higher momenta, but since we are interested in the early stage of the process, we consider only first order side-modes to be populated.

\subsection{The Model}
\label{secIIa}
Neglecting any coupling {\it between} counterpropagating photonic endfire modes, 
the dynamics in the two halves of the elongated system become  
independent
. We can thus focus on the endfire modes 
with ${\bf k}\approx +k{\bf e}_z$ and consequently on the two sets of atomic side-modes 
(with central momenta $\hbar{\bf q}=\pm\hbar(k_l{\bf e}_x - k {\bf e}_z)$) 
that are coupled to them. Due to the strong confinement along the $x$- and $y$-axis, 
we can assume the transverse profiles of  the matter field $\psi_\perp(x,y)$  to be well 
described by a classical function, independent of the $z$-coordinate \cite{ZobNikPRA}. 
Assuming a Fresnel number close to unity for the electromagnetic fields, we can apply the 
same approximation for the transverse part of the radiation field $u_\perp(x,y)$, effectively 
reducing the problem to one dimension.

We expand the field operators as
\numparts\label{expansions}
\begin{eqnarray}
\hat{\psi}({\bf x},t)&=&\psi_\perp(x,y)\sum_{j=-1}^1\hat{\psi}_{j}(z,t)\rme^{{\mathrm i}j(k_lx-kz)-{\textrm i}\omega_jt}\label{expandpsi}\\
\label{expandE}
\hat{\bf E}^{(+)}({\bf x},t) &=&\frac{{\mathcal E}_0}{2}{\bf e}_y \rme^{{\mathrm i}(k_lx-\omega_l t)} + u_\perp(x,y)\hat{E}_+^{(+)}(z,t){\bf e}_y,
\end{eqnarray}
\endnumparts
where $\omega_{\pm 1}=\hbar(k_l^2+k^2)/2M$ with $M$ the atomic mass and $\omega_0=0$. The matter-wave operator is split up in three parts, describing the two side-modes ($j=\pm 1$) and the BEC at rest ($j=0$), 
\numparts\label{modeexp}
\be
\hat{\psi}_{j}(z,t)=\rme^{{\mathrm i}\omega_{j}t}\sum_{p\in\Delta_{0}}\frac{\rme^{\mathrm i p z}}{\sqrt{L}} {\hat c}_{-j k+p}(t),\label{psipm}
\ee
where $\Delta_0$ is the interval $(-k/2,k/2)$ in $k$-space, $L$ is the length of the BEC and $\hat{c}_p$ annihilates an atom with momentum $\hbar p$. Since the BEC at rest remains practically undepleted it can be treated as a time independent classical function and hence we can set $\hat{\psi}_{0}(z,t)\equiv\psi_0(z)$. Similarly, we expand the endfire mode operator as
\bea
\hat{E}_+^{(+)}({\bf z},t) &=& \mathrm i \rme^{\mathrm i(kz-\omega t)} \sum_{p\in\Delta_{0}} \sqrt{\frac{\hbar \omega_{k+p}}{2\varepsilon_0 }}  \frac 1{\sqrt L} \rme^{\mathrm ipz}\rme^{\mathrm i\omega t} \hat a_{k+p}(t)\nonumber\\
&\approx& \sqrt{\frac{\hbar \omega}{2\varepsilon_0}}\rme^{\mathrm i(kz-\omega t)}\hat e_+(z,t)\label{exp_Ep},
\eea
\endnumparts
where $\hat{a}_p$ is the photon annihilation operator. The frequencies $\omega_{k+p}$ are approximated by $\omega=k/c$, the frequency of the scattered photons, and can therefore be taken out of the sum. This approximation is justified by the fact that dominant contributions to the sum come from momenta of order $1/L$, which is several orders of magnitude smaller than $k$.

The one-dimensional field operators satisfy usual boson commutation relations
\numparts
\bea\label{commrel}
\left[\hat{\psi}_i(z_1,t),\hat{\psi}_j^\dag(z_2,t)\right]&=&\delta_{ij}\delta_\Delta(z_1-z_2),\label{commrelpsi}\\
\left[\hat{e}_{+}(z_1,t),\hat{e}_{+}^\dag(z_2,t)\right]&=&\delta_\Delta(z_1-z_2),\label{commrelExi}\\
\left[\hat{e}_{+}(z,t_1),\hat{e}_{+}^\dag(z,t_2)\right]&=&\frac{1}{c}\delta_\Delta(t_1-t_2),\label{commrelEtau}
\eea
\endnumparts
where $\delta_{ij}$ denotes the Kronecker delta and $\delta_{\Delta}(z)$ is a delta-like distribution with width of order $1/k$. Since the fields vary slowly on length scales of the order of $1/k$, we will approximate the $\delta_{\Delta}(z)$ by actual delta functions.
Inserting expansions (\ref{expansions}) in the Maxwell-Schr\"odinger equations for the coupled matter-wave and electric fields and applying the slowly-varying-envelope approximation, we find the equations of motion for the operators 
\numparts\label{eom}
\begin{eqnarray}
\frac{\partial \hat{\psi}_{+1}^\dag(\xi,\tau)}{\partial\tau}&=&
{\rm i}\kappa \hat{e}_+(\xi,\tau) \psi_{0}^*(\xi)
,\label{eom1-1}\\
\frac{\partial \hat{\psi}_{-1}^\prime(\xi,\tau)}{\partial\tau}&=&
-{\rm i}\kappa \hat{e}_+(\xi,\tau)\psi_{0}(\xi) 
-2{\rm i} \hat{\psi}_{-1}^\prime(\xi,\tau)
,\label{eom-11}\\
\frac{\partial \hat{e}_+(\xi,\tau)}{\partial\tau}+
\chi \frac{\partial \hat{e}_+(\xi,\tau)}{\partial\xi}
&=&-{\rm i}\left [\kappa\psi_{0}(\xi)
\hat{\psi}_{+1}^\dag(\xi,\tau)\right.\nonumber\\
&&\left.\quad+
\kappa\hat{\psi}_{-1}^\prime(\xi,\tau)
\psi_{0}^*(\xi)\right ].
\label{eome+}
\end{eqnarray}
\endnumparts
Here we have defined $\hat{\psi}_{-1}^\prime(\xi,\tau)=\rme^{-2{\mathrm i}\tau}\hat{\psi}_{-1}(\xi,\tau)$ and rescaled length  and time to dimensionless units 
\bea
\xi=k_lz, \qquad \tau=2\omega_rt,
\eea
where $\omega_r=\hbar k_l^2/2M$. Accordingly, the fields are rescaled as
\be
\hat{e}_+(\xi,\tau)\equiv\frac{1}{\sqrt{k_l}}\hat{e}_+(z,t)\qquad\hat{\psi}_{j}(\xi,\tau)\equiv\frac{1}{\sqrt{k_l}}\hat{\psi}_{j}(z,t)\nonumber,
\ee
and the speed of light becomes $\chi\equiv\frac{ck}{2\omega_r}$.
The effective 1D-coupling is given by $\kappa=g\sqrt{k_lL}/(2\omega_r)$ with
\be
g=\frac{|{\bf d}\cdot{\bf e}_y|^2{\mathcal E}_0}{\hbar^2\delta}\sqrt{\frac{\hbar\omega}{2\epsilon_0L}}\int{\mathrm d}x{\mathrm d}yu_\perp(x,y)\psi_\perp^2(x,y).\nonumber
\ee
For a detailed derivation of (\ref{eom}) see \cite{BuNikZobLam}. 

Backwards recoiling atoms are a particular feature of superradiant Rayleigh scattering off condensates. 
The physical process underlying the backwards modes violates energy conservation by an amount 
$\Delta E\simeq 4 \hbar\omega_{r}$. Thus, according to Heisenberg uncertainty principle, it can take place 
only for times shorter than a critical time $\hbar/\Delta E$ which, in our units, is given by \cite{InoChiSta99},
\be 
\tau_c=0.5.
\ee
For such short pulses, and for sufficiently high power one typically 
observes an X-shaped pattern for the distribution of the atomic side-modes with the initial BEC in the center and the 
recoiling atoms moving both in and against the direction of the applied laser pulse 
({\it Kapitza-Dirac} or {\it strong-pulse regime}). On the other hand, 
for weaker pulses with duration longer than $\tau_{c}$, 
the distribution of the side-modes exhibits a fan pattern, involving mainly forward recoiling atoms 
({\it Bragg} or {\it weak pulse regime}). 
If we neglect the atomic backwards side-mode altogether, (\ref{eom}) become formally 
equivalent to descriptions of ``conventional'' superradiance in excited atomic gases \cite{Dicke}. 
 
\subsection{Solutions to the equations of motion}
\label{secIIb}
We can use the Laplace transform to find exact solutions to the system (\ref{eom}) in terms of the operators evaluated at the boundary of their domain -- i.e. at $\xi=0$ and $\tau>0$ 
and at
$\xi>0$ and $\tau=0$. More details on this procedure are given in \cite{BuNikZobLam}. In the following we are interested in the radiation field only, whose 
time evolution
 is given by
\bea
\hat{e}_+(\xi,\tau)&=&
\int_0^\tau \rmd\tau^\prime \hat{e}_+(0,\tau^\prime ) F_{0,0}(\gamma_{\xi,0},\tau-\tau^\prime)\nonumber\\
&&-\frac{{\rm i}\kappa}{\chi}
\int_{0}^{\xi}\rmd\xi^\prime
\left[
\psi_{0}(\xi^\prime)
\hat{\psi}_{+1}^\dag(\xi^\prime,0) F_{1,0}(\gamma_{\xi,\xi'},\tau)\right.\nonumber\\
&&\left.+\psi_{0}^*(\xi^\prime)
\hat{\psi}_{-1}^\prime(\xi^\prime,0) F_{0,1}(\gamma_{\xi,\xi'},\tau)\right ],
\label{solutions}
\eea
where we have neglected retardation effects and introduced
\be
\gamma_{\xi,\xi'}=\frac{\kappa^2}{\chi}[\rho(\xi)-\rho(\xi')]\nonumber,
\ee
with $\rho(\xi)=\int_0^\xi \rmd\xi' |\psi_{0}(\xi')|^2$. The functions $F_{\mu,\nu}(u,v)$ are defined as
\be
F_{\mu,\nu}(u,v)={\mathcal L}_{p\to v}^{-1}\left\{\frac{\rme^{u/p}\rme^{-u/(p+2{\mathrm i})}}{p^\mu(p+2{\mathrm i})^\nu}\right\}\nonumber,
\ee
where ${\mathcal L}^{-1}_{p\to v}$ denotes the inverse Laplace transform. 
Explicit expressions for the functions $F_{\mu,\nu}(u,v)$ appearing in (\ref{solutions}) are given in the appendix. They are combinations of Bessel functions and integrals thereof. 
For the numerical calculations, we assumed the BEC to consist of $N=10^6$ Thomas-Fermi distributed ${}^{87}\textrm{Rb}$ atoms, such that $\psi_0=\sqrt{\Theta(z)N6(Lz-z^2)/L^3}$ with $L=130\mu{\textrm m}$. For the incoming laser we chose a rectangular profile and a wavenumber $k_l=8.05\times10^6{\textrm m}^{-1}$, which results in a dimensionless length of the BEC of $\Lambda\sim 1000$. Coupling strengths are conveniently expressed in terms of the {\it superradiant gain} $\Gamma=\kappa^2N/\chi$, whose value separates the two regimes identified by experimental observations of superradiance from condensates 
\cite{SRExp}. Typically, the weak coupling regime is characterized by $g \sim 10^5{\textrm s}^{-1}$ and $\Gamma\sim1$, while $g\sim10^6{\textrm s}^{-1}$ and $\Gamma\gg 1$ in the strong coupling regime. In our calculations, we chose $\Gamma=1$ and $\Gamma=100$ for the two regimes. 
 
\subsection{Averaging over semiclassical trajectories}
\label{secIIc}

The presented model is closely related to the mean-field (MF) model presented in \cite{ZobNikPRA}. In fact, we arrive at the MF equations with the same approximations by replacing all operators in (\ref{eom}) by their expectation value. Due to the generality of the Laplace transform, the same replacement in (\ref{solutions}) gives the solutions to the MF model. A detailed comparison between the two models is given in \cite{BuNikZobLam}. One of the problems of the MF model is the fact that it cannot describe the spontaneous onset of the superradiance process. In order to start the evolution, one has to ``seed'' one of the modes with some initial values.
In the context of conventional superradiance, Haake {\it et al.} introduced the idea of averaging over many semiclassical ``trajectories'' to obtain quantum results \cite{Haake}. The MF equations are initially  seeded with random variables according to a particular distribution. To obtain a particular quantum expectation value, one has to average over various solutions for the corresponding semiclassical quantity. We have investigated the extension of this idea to superradiant Rayleigh scattering off BECs where, in contrast to conventional superradiance,  backwards recoiling atoms are also present. 

To see how this works, let us assume we want to calculate the normal-ordered $n$th order correlation function of the emitted light.
\be
\langle\left[\hat{e}_+^\dag(\xi,\tau)\right]^n\left[\hat{e}_{+}(\xi',\tau')\right]^n\rangle.
\label{egcorrfunc}
\ee
From (\ref{solutions}) and using the commutation relations as well as the fact that our initial state is the vacuum for all modes, we find the only non-vanishing expectation value involved to be 
\be
\langle\hat\psi_{+1}(\xi^{(1)},0)\ldots\hat\psi_{+1}(\xi^{(n)},0)\hat\psi_{+1}^\dag(\xi^{(n+1)},0)\ldots\hat\psi_{+1}^\dag(\xi^{(2n)},0)\rangle,
\label{normprodexpvalue}
\ee
where the $\xi^{(j)}$ are integrated from $0$ to $\xi$ for $j=1,\ldots,n$ and from $0$ to $\xi'$ for $j=n+1,\ldots,2n$. Using the commutation relations, correlation (\ref{normprodexpvalue}) reads
\be
\sum_\pi\prod_{j=1}^n\delta(\xi^{(j)}-\xi^{(n+\pi(j))}),
\label{contractions}
\ee
where $\delta$ denotes the Dirac delta function and the sum runs over all permutations $\pi$ of order $n$. 

Let us now seed the semiclassical model with \cite{BuNikZobLam}
\numparts
\label{seeding}
\bea
&&\psi_{+1}(\xi,0)=C_\xi,\\
&&\psi_{-1}(\xi,0)=0,\\
&&e_{+}(\xi,0)=0,
\eea
\endnumparts
where $C_\xi$ is a random,  normally distributed complex variable, with zero mean and variance 
$1/\sqrt{\Delta\xi}$, with $\Delta\xi$ the spatial step of a numerical implementation\footnote{In our case 400 spatial gridpoints}. 
The average of the product $\psi_{+1}(\xi^{(1)},0)\ldots\psi_{+1}(\xi^{(n)},0)\psi_{+1}^*(\xi^{(n+1)},0)\ldots\psi_{+1}^*(\xi^{(2n)},0)$ over many trajectories (seedings) will effectively converge towards (\ref{contractions}). 
Due to the formal equivalence of the semiclassical solutions to the quantum ones, the average of the product 
${e_{+}^*(\xi_1,\tau)}^n e_{+}(\xi_2,\tau)^n$ will consequently converge towards the quantum expectation value 
(\ref{egcorrfunc}). To find correlation functions of other operators, other seedings have to be used, 
which can be found in an analogous way and are summarized in \cite{BuNikZobLam}. 

\begin{figure}
\begin{center}
\includegraphics{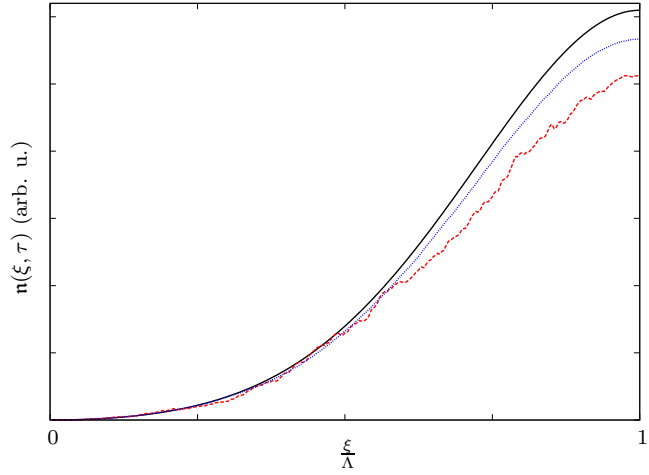}
\end{center}
\caption{
(Colour online) Example of the photon density within the BEC volume obtained through averaging over randomly seeded MF solutions.  
The quantum solution is shown in solid black, while averages over 20 and 2000 trajectories are shown in red dashed and 
blue dotted lines. Parameters: $\Gamma=1$, and $\tau=2$.}
\label{averagefig}
\end{figure}

As far as densities and populations are concerned, the convergence of the averaging procedure towards the quantum 
solution is fairly fast. 
In figure \ref{averagefig} we plot the photon density
\be
\mathfrak{n}(\xi,\tau)=\langle\hat{e}_+^\dag(\xi,\tau)\hat{e}_+(\xi,\tau)\rangle
\label{density}
\ee
averaged over different numbers of trajectories compared to the quantum result.
Typically, one has to average over a 
couple of thousand trajectories, to obtain well converged, smooth density profiles. 
Clearly, the averaging procedure is also applicable to correlations of higher order. 
One can calculate any normal ordered correlation function of the system by averaging over a sufficiently 
large ensemble of MF trajectories, provided that the operators involved in the correlation function have the 
same seeding requirements. The convergence, however, becomes rapidly slower with every order added, 
such that higher order correlations will require a larger number of trajectories. 
Higher order correlations are related to higher order statistical moments of a particular 
observable, such as the variance or skew of a particular distribution. 

\section{Passage-Time Statistics}
\label{secIII}
We proceed to apply the technique of averaging over semiclassical trajectories to the study of passage-time statistics. First, we will find a simple probability density function which matches the distributions. Following this, we will investigate the behaviour of the distributions with respect to scaling of the two most important parameters of such an experiment and find invariant quantities for both cases.
\subsection{Distribution of Passage-Times}
In view of Sec. \ref{secIIc}, a single semiclassical trajectory represents a single 
experiment on superradiance off a given BEC \cite{BuNikZobLam,GroHar82,Haake}. The process is initially driven by quantum 
fluctuations; within the MF model this is simulated by seeding $\psi_{+1}$ 
with Gaussian white noise. After the first spontaneous scattering events, bosonic stimulation 
leads to a rapid growth of the emitted radiation and of the side-mode populations. 
The signature of the quantum fluctuations which started the process, however,  is imprinted onto the superradiant light pulses.

One way to investigate quantum fluctuations in this context, is by studying the {\em first passage-time statistics}, 
which pertain to the distribution of times at which the intensity of the SR pulse reaches a certain reference intensity. 
Equivalently, one may look at the times for which the number of photons that have left the condensate volume, 
given by 
\be
\mathscr{N}(\tau)=\chi\int_0^\tau\mathrm{d}\tau'\mathfrak{n}(\Lambda,\tau'),  
\label{photout}
\ee
reaches a certain threshold $\mathscr{N}_{\rm th}$. 
Performing a large number of semiclassical simulations, each one seeded with a 
new random $\psi_{+1}$, and following the evolution of $\mathscr{N}(\tau)$ for every trajectory, 
one can record the corresponding passage-times $T$. 
The quantum fluctuations responsible for the start-up of the system are translated into fluctuations of the passage times, whose statistics can provide  
information about the onset of the phenomenon.
Moreover, due to the rapid growth of the mode populations and the deterministic evolution of the system, the passage-time statistics are expected to also reflect the delay-time statistics,  
which refer to the times at which the first maximum of the SR pulse occurs.

Based on these ideas, Haake {\it et. al.} investigated the statistics of passage- (and delay) times 
of SR light pulses obtained from a conventional superradiant system \cite{Haake}. 
A semi-analytic expression was given for the distribution of the passage-times, which 
was found to be well approximated by the Gumbel probability density function (PDF) \cite{GroHar82} 
\be
{\mathcal P}_{\mathrm G}(T;\mu,\lambda)=\lambda^{-1}\rme^{-\frac{T-\mu}{\lambda}}\rme^{-\exp(-\frac{T-\mu}{\lambda})}.
\ee
Their predictions about the standard deviation of the passage-times were found to be in 
good agreement with experimental data \cite{PRAVreWed}.

\begin{figure}
\begin{center}
\includegraphics[]{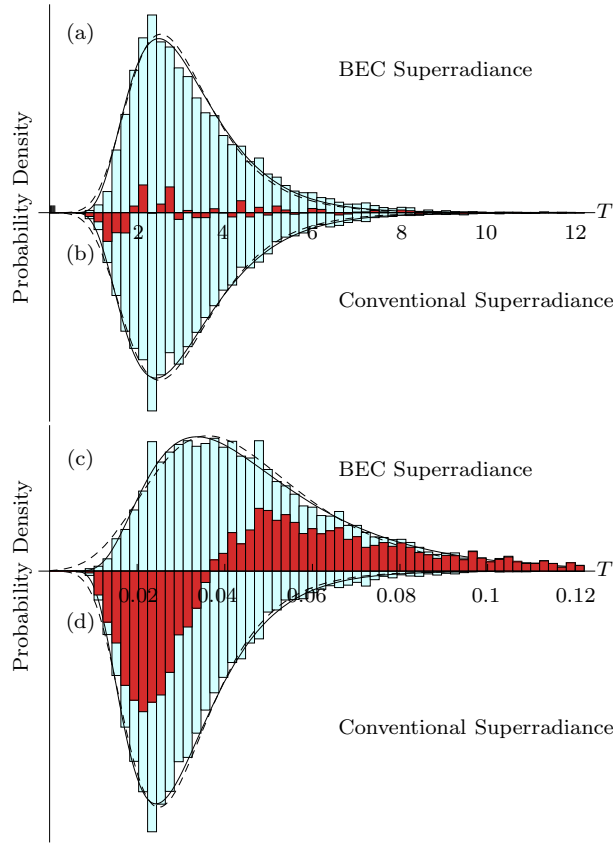}
\caption{(Colour online) Passage-time statistics of light pulses in superradiance from BECs. Histograms (a) and (c) show the distribution of the times at which 
the number of photons that have left the BEC reaches the threshold $\mathscr N_{\mathrm{th}}=10$,
for (a) $\Gamma=1$, and (c) $\Gamma=100$. The corresponding 
histograms with neglected backwards recoiling atoms (equivalent to 
conventional superradiance), are plotted with inverted ordinate (b,d).
Dark(red) histograms show the difference between the two cases, in the 
two regimes. The curves give least square fits with inverse Gaussian 
(solid) and Gumbel (dashed) PDFs.}
\label{histograms}
\end{center}
\end{figure}

In contrast to conventional superradiant systems, in superradiance off BECs we also have the presence of 
backwards atomic side-modes. This is an essential feature of our system and can influence statistical properties of the scattered light, such as the the first- and second-order correlation functions \cite{BuNikZobLam}. 
Here, we will discuss its influence on the passage-time statistics. 
To this end, we performed a number of numerical simulations using the semiclassical equations of motion. 
For each trajectory, the equations were seeded in accordance with (\ref{seeding}), 
and we followed the time-evolution of the number of outgoing photons $\mathscr N(\tau)$. 
As soon as the threshold  ${\mathscr N}_{\rm th}$ was reached, we recorded the corresponding time 
$\tau=T$, and started a new simulation with new random seeding. 
The histograms depicted in figure \ref{histograms} describe the distribution 
of the recorded times after a large number of simulations.  
 
\begin{figure}
\begin{center}
\includegraphics[]{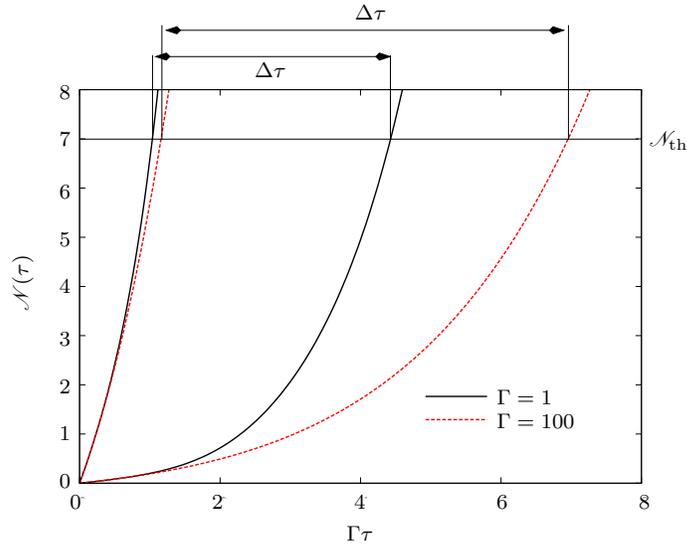}
\caption{(Colour online) Evolution of the number of photons in the SR light pulse for two different seeds, as a function of scaled time in the strong and weak coupling regime. The vertical lines indicate the wider spread due to the slower growth in the strong coupling regime.
}
\label{seedplot}
\end{center}
\end{figure}

In the weak pulse regime, the distributions are narrower 
(i.e., the variance of the distributions is smaller) than in the strong pulse regime. 
An intuitive understanding for this broadening can be obtained by recalling that in the strong pulse regime, 
the growth of the photonic endfire modes is slowed down by the presence of backwards recoiling atoms, as each backscattered atom removes a photon from the endfire modes. 
Therefore,  with respect to the scaled time $\Gamma\tau$, the spread of solutions for ${\mathscr N}(\tau)$ 
corresponding to different seedings, is larger in the strong than in the weak pulse regime. 
This is illustrated in figure \ref{seedplot}, where the evolution of ${\mathscr N}(\tau)$ is plotted 
for two different seedings in the strong- and the weak-pulse regimes. 

All the distributions exhibit positive skew, which limits the PDFs 
appropriate to model the data. 
Several two-parameter, positively skewed PDFs were fitted to the data according to a least-square 
fit  and assessed by calculating the residual errors as well as comparing mean, variance and skewness of the distributions 
to the ones of the data. While the Gumbel PDF of conventional superradiance gives a very 
satisfying fit, it is for all parameter values studied inferior\footnote{The assessment of fitted PDFs was performed by calculating the sum of residual errors as well as by checking the agreement of the PDFs mean, variance, skew and kurtosis with the ones of the raw data.} to a fit with the inverse Gaussian PDF
\be\mathcal P_{\mathrm{IG}}(T;\mu,\lambda)=\sqrt{\frac{\lambda}{2\pi T^3}}\rme^{-\lambda\frac{(T-\mu)^2}{2\mu^2 T}}\label{invgaussdist},
\ee
which, as a further advantage, does not support negative $T$. 
If a random variable $X$ is distributed according to (\ref{invgaussdist}), 
the mean and variance of $X$ are given by
\be
\bar X=\mu, \quad \mathrm{var}(X)=\mu^3/\lambda.
\label{meanvariance}
\ee
In figure \ref{histograms} we show two fits to the numerical data, 
corresponding to the Gumbel and the inverse Gaussian PDFs. 

We now turn to the comparison of the passage-time distributions for superradiance off BECs, to the ones 
obtained for conventional superradiance. As discussed in section \ref{secIIa}, the latter system can be simulated within 
our model by neglecting the backwards atomic side-modes in  (\ref{eom}).  
The corresponding distributions of the recorded passage-times are depicted in figures \ref{histograms} (b) and (d). 
From the difference of the two histograms, also plotted in figure \ref{histograms}, we see that in the weak pulse regime $(\Gamma=1)$, 
the two distributions are virtually the same [compare figures \ref{histograms} (a) and (b)]. For weak couplings, 
the number of backwards recoiled atoms is not big enough to change dramatically the passage time statistics. 
In the strong pulse regime, however, the distribution for conventional superradiance is narrower than the 
one for superradiance off BECs, while it is also shifted towards smaller passage times [compare figures \ref{histograms} (c) and (d)]. 
Finally, in the weak pulse regime, fits with either of the aforementioned 
two PDFs agree very well with the distributions of the numerical data 
[see figure \ref{histograms} (a) and (b)]. On the contrary, in the strong-pulse regime, and if backwards 
recoiled atoms are present, we find noticeable discrepancies between the two fits, particularly for small $T$. They become more prominent 
for small thresholds $\mathscr N_{\mathrm{th}}$.

\subsection{Scaling with coupling strength}
We have seen that our equations of motion can describe conventional superradiance if we set all the couplings  
to backward atomic side-modes to zero. The resulting reduced set of equations is also expected to be a good approximation in the 
weak-coupling regime of superradiance off condensates, provided that the time scales of interest are such that 
the backward atomic side-modes are scarcely populated. Assuming furthermore a homogeneous condensate, analytic solutions for the operators can be found, which moreover are a function of the scaled time $\Gamma\tau$ rather than just the time alone \cite{BuNikZobLam}. 
If one multiplies inverse-Gaussian distributed random numbers with a constant factor, the resulting numbers are again inverse-Gaussian distributed, with the mean $\mu$ and the shape parameter $\lambda$ rescaled by the same number. Hence the scaling behaviour is also  reflected in the distributions, which can be seen by comparing the distributions of figure \ref{histograms} (b) and (d), where a scaling factor or 100 is clear.
We can therefore extend our previous observation on the scaling behaviour of the passage-time 
distributions for conventional superradiance, to the context of superradiance off condensates and define a more general ``scaled'' version of the inverse Gaussian PDF, 
which describes accurately passage-time distributions for various couplings within 
the weak-coupling regime, 
\be
\mathcal P_{\mathrm{WP}}(\tau;\mu,\lambda)=\mathcal P_{\mathrm{IG}}(\tau,\mu_{\mathrm{WP}}/\Gamma,\lambda_{\mathrm{WP}}/\Gamma).
\label{scaling}
\ee
For any $\mathscr{N}_{\mathrm{th}}$, the two open parameters $\mu_{\mathrm{WP}}$ and $\lambda_{\mathrm{WP}}$ are constants characterizing the weak-pulse regime. For $\mathcal{N}_{\mathrm{th}}=10$, their value was estimated to $\mu_{\mathrm{WP}}\approx 3.2$ and $\lambda_{\mathrm{WP}}\approx 16$. 
In the case of conventional superradiance, it is known that the mean of the passage times scales as a constant 
divided by the atomic decay rate, while their variance scales as another constant divided by the decay rate 
squared \cite{GroHar82}. In our model, the quantity corresponding to the decay rate is the superradiant gain 
$\Gamma$, which is directly proportional to the superradiant scattering rate. From (\ref{meanvariance}) 
and (\ref{scaling}) we therefore find that the inverse Gaussian distribution recovers the same scaling laws as found in \cite{GroHar82}.

\begin{figure}
\begin{center}
\includegraphics[]{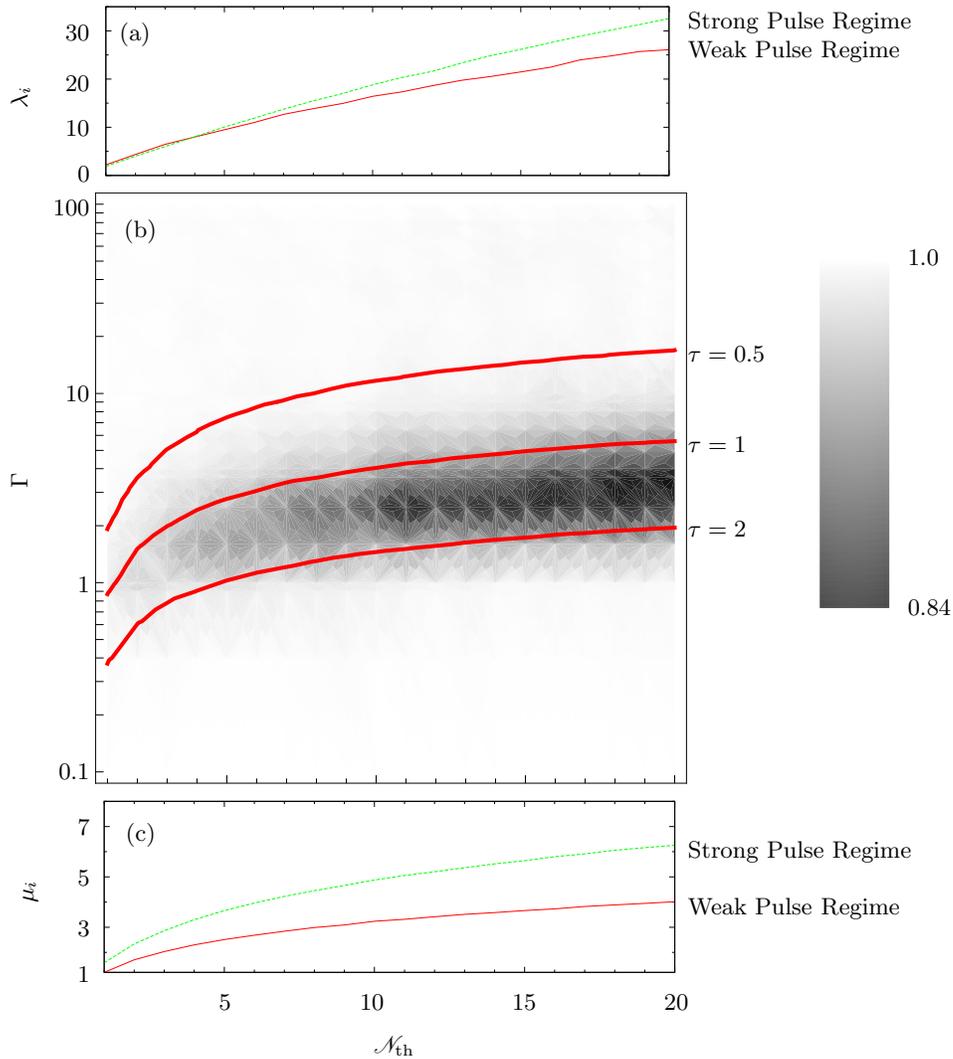}
\caption{(Colour online) (a) Shape parameters of the inverse gaussian PDFs as a function of photon threshold $\mathscr{N}_\mathrm{th}$. (b) Density plot of the overlap (\ref{overlapdef}), for various couplings $\Gamma$ and thresholds $\mathscr N_{\mathrm{th}}$. White corresponds to a value of unity. Note the logarithmic scale for $\Gamma$. The contour lines indicate the points for which the expectation time for $T$ is equal to 0.5,1 and 2 respectively. Each histogram contains $10^4$ trajectories. (c) Means $\mu$ of the scaled passage times as a function of photon threshold $\mathscr{N}_\mathrm{th}$.}
\label{scalingfig}
\end{center}
\end{figure}

If we increase the coupling strength and leave the weak-coupling regime, we lose this scaling behaviour, 
as can be seen by comparing the histograms (a) and (c) in figure \ref{histograms}. 
Surprisingly, however, we recover (\ref{scaling}) deep within the strong pulse regime, albeit with new 
constant parameters, which for $\mathscr{N}_{\mathrm{th}}=10$ are given by $\mu_{\mathrm{SP}}\approx5$ and $\lambda_{\mathrm{SP}}\approx 20$. 
To gain further insight into the change of the scaling behaviour of the passage-time distribution 
with the coupling strength, we calculated numerically the distributions ${\cal H}(\Gamma T,\mathscr{N}_{\rm th})$, 
of the rescaled passage-times $\Gamma T$, for various values of $\Gamma$ with $\Gamma_{\rm low}\leq \Gamma\leq \Gamma_{\rm high}$,
and various thresholds $\mathscr{N}_{\rm th}$. The upper and lower values of $\Gamma$ were chosen as 
$\Gamma_{\rm low}=0.1$ and $\Gamma_{\rm high}=100$, so that both regimes are covered.     
Subsequently, for a given $\mathscr{N}_{\rm th}$, we calculated the normalized overlap 
\be\label{overlapdef}
\max_{i\in\{\rm low,up\}}\bigg\{\frac{\left[\int_0^\infty\mathcal H(\Gamma T,\mathscr{N}_{\rm th})\mathcal H(\Gamma_{i}T,\mathscr{N}_{\rm th})\mathrm{d}T\right]^2}{\int_0^\infty\mathcal H(\Gamma T,\mathscr{N}_{\rm th})^2\mathrm{d}T \int_0^\infty\mathcal H(\Gamma_{i} T,\mathscr{N}_{\rm th})^2\mathrm{d}T}\bigg\},
\ee
and the corresponding density plot is depicted in figure \ref{scalingfig}. While the parameters $\mu_{\mathrm{WP/SP}}$ and $\lambda_{\mathrm{WP/SP}}$ are independent of the coupling, they are nontrivial functions of the photon threshold $\mathscr{N}_\mathrm{th}$, and their numerical dependence is also depicted in figures $\ref{scalingfig}$ (a,c).
Quite obviously, the scaling law within the two regimes is very well obeyed by the system. 
We can therefore conclude that in the early stages of superradiant scattering off condensates, 
the distribution of passage times in either regime is universally well modelled by the 
inverse Gaussian PDF and two constant parameters.  The transition between the two 
scaling behaviours begins in the region, where the parameters $\mathscr{N}_{\rm th}$ 
and $\Gamma$ match up in a way such that $\bar{T}\approx \tau_\mathrm{c}$. The distributions in this transient regime do 
not differ qualitatively from the ones presented in figure \ref{histograms}, but one cannot define a 
universal scaling law for them. 
As can be seen in figure \ref{scalingfig} (b), however, this transient regime (dark region) in the parameter space 
is rather small compared to the ones where the system is unambiguously found in either of the two regimes (white regions). 
As a final remark, it should be noted that for any threshold $\mathscr N_\mathrm{th}$, the ratio of mean and shape parameter of the resulting distribution is independent of $\Gamma$,  which by virtue of (\ref{meanvariance}) implies that the ratio of the variance and the squared mean of passage times is independent of $\Gamma$. 

\subsection{Scaling with threshold photon number}
\begin{figure}
\begin{center}
\includegraphics[]{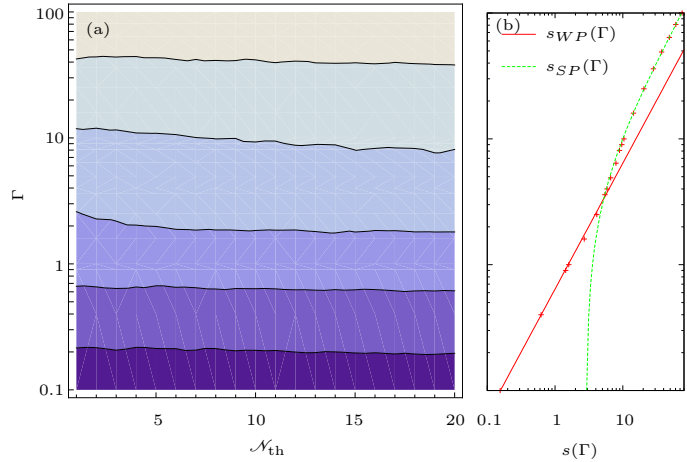}
\caption{(Colour online) (a) Contour plot of the value of $s$ as defined in (\ref{def_s}). (b) The parameter $s$, averaged over $\mathscr{N}_\mathrm{th}$, as a function of the coupling $\Gamma$ (crosses) and the fits 
(24) in the weak and strong pulse regimes (lines). For every threshold and coupling $10^4$ trajectories were evolved.
}
\label{brownmotfig}
\end{center}
\end{figure}

Having investigated the behaviour of the passage time distributions for varying $\Gamma$, it is natural to ask how they behave with a constant coupling but varying the other significant parameter of the system, the photon threshold $\mathscr N_{\mathrm{th}}$.  For any choice of $\Gamma$, we find  again a simple combination of shape parameter and mean, which remains constant independent of $\mathscr N_{\mathrm{th}}$. It is the ratio 
\be\label{def_s}
s=\frac{\lambda}{\mu^2}=\frac{\bar T}{\mathrm{var}(T)},
\ee
where the second equality is a simple consequence of (\ref{meanvariance}). Figure \ref{brownmotfig} (a) depicts a contour plot of the value of $s$ for different couplings and photon thresholds, and reveals the independence of this value of the coupling $\Gamma$. Since these results are based on the analysis of a large ensemble of random trajectories, a certain variation is to be expected, and the values of $s$ as a function of $\Gamma$ are obtained by averaging $s$ over $\mathscr N_\mathrm{th}$. Moreover, for each regime, the quantity $s$ has a very simple linear dependence on the coupling $\Gamma$, which is illustrated in figure (\ref{brownmotfig}) (b). The functions plotted are the fits
\numparts
\begin{eqnarray}\label{sofgamma}
s_\textrm{WP}(\Gamma)&=1.56\Gamma\qquad\qquad&\textrm{(weak pulse regime)}\\
s_\textrm{SP}(\Gamma)&=0.71\Gamma+2.86&\textrm{(strong pulse regime)}.
\end{eqnarray}
\endnumparts
Note that the equation for the strong pulse regimes is valid for large values of $\Gamma$, such that the offset becomes negligable and $s_\textrm{SP}$ is directly proportional to $\Gamma$. Hence this relation is consistent with equation \ref{scaling}.
Moreover, also in this case the transition between the two fits 
(24), takes place in a regime where the parameters result in times around critical time $\tau_\mathrm{c}=0.5$.

Random variables distributed according to the inverse Gaussian PDF and with this very ratio as a constant, also appear in different contexts. Consider the stochastic process  
\numparts\label{stochproc}
\bea
X_0&=&0\\
X_t&=&\nu t+\sigma B_t,
\eea
\endnumparts
where $B_t$ is the standard Brownian motion and $\nu>0$ is a positive, constant drift. The times at which $X_t$ crosses a certain limit $\alpha>0$ are distributed according to the inverse Gaussian distribution \cite{ChhFolk}
\bea
T_\alpha&=&\inf\{0<t|X_t=\alpha\}\nonumber\\
&\sim&\mathcal P_{\mathrm{IG}}\left(T_\alpha;\mu=\frac{\alpha}{\nu},\lambda=\frac{\alpha^2}{\sigma^2}\right).
\label{firstpass}
\eea
The ratio of shape parameter and squared mean is independent of $\alpha$, which is analogous to $s$ being independent of the photon threshold $\mathscr{N}_\mathrm{th}$ and we can thus make a connection between Brownian motion and superradiance off BECs. In particular, we can construct a stochastic system analogous to (\ref{stochproc}), but in terms of $\bar T$ and $s$. Note that the particular choice of corresponding parameters is not unique. From  (\ref{firstpass}) we find that $\alpha/\nu$ must correspond to the quantum expectation value $\bar T$, while 
\be
s=\frac{\nu^2}{\sigma^2}.
\ee
We can therefore set the drift $\nu$ to unity and find $\bar T = \alpha$ and $\sigma=1/\sqrt{s}$. Hence,  the times the stochastic system
\bea
X_0&=&0,\\
X_\tau&=&\tau+\frac{1}{\sqrt{s}}B_\tau,
\eea 
will cross $X_\tau=\bar T$ will reproduce the distribution of passage-times for 
a superradiant BEC emitting $\mathscr N_{\mathrm{th}}$ photons.
This result by no means implies that superradiance is in fact a physical realization of Brownian motion. In contrast to Brownian motion, superradiance off condensates is a \emph{deterministic} evolution from random initial conditions. After its initial stage, the system can be described by deterministic equations. This is never the case for Brownian motion.

\section{Conclusion}
\label{conclusion}
We have used random seedings to investigate the passage-time statistics 
of superradiant light pulses emitted from an elongated atomic 
condensate. In general, the distributions of the passage times are very 
well approximated by the inverse Gaussian probability density function. 
The distributions in the Kapitza-Dirac regime, however, appear wider 
than in the Bragg regime, while they are also shifted towards larger
passage times.  This is due to the presence of backwards recoiling atoms 
in the system, which in the Kapitza-Dirac regime lead to a slow growth 
of end-fire modes. For both regimes the mean of the distributions is 
inversely proportional to the superradiant gain $\Gamma$, whereas 
their variances scale with $\Gamma^{-2}$. The corresponding 
proportionality constants are different in the two regimes and for each regime, the ratio of shape parameter $\lambda$ and mean $\mu$ is independent of $\Gamma$.
For a small region in parameter space, the system is not in 
either of the two regimes but finds itself in a transient one, where the 
distribution of the passage times changes smoothly between the two 
reference (outer most) distributions. This transient region is 
characterized by timescales around the critical time, at which energy 
conservation suppresses backwards recoiling atoms.
For constant coupling and varying threshold, the ratio $\lambda/\mu^2$ is constant. The latter behaviour leads to a phenomenological correspondence between superradiance from BECs and Brownian motion with drift. We focused at the very early stage of the phenomenon, but due to the deterministic evolution of the system the observed behaviour can be expected to hold for later times and consequently the distributions are expected to also describe the delay-times of a superradiant pulse from a condensate. 
Finally we want to stress that the presented results are accessible to experimental verification. Similar experimental studies have been performed for conventional superradiance \cite{GroHar82,PRAVreWed}.

\begin{appendix}
\section{Inverse Laplace Transforms}
Using elementary techniques for Laplace inversion, one finds
\numparts
\bea
F_{1,0}(y,z)&=&
I_0\left ( 2\sqrt{y z} \right )\Theta( z)\nonumber\\
&&-\Theta( z)\sqrt{y}\int_0^{ z} \rmd z^\prime \frac{\rme^{-{\rm
i}2z^\prime}}{\sqrt{z^\prime}}
I_0\left [ 2\sqrt{y(z-z^\prime)} \right ]
J_1\left ( 2\sqrt{y z^\prime} \right )\\
F_{0,1}(y,z)&=&
\rme^{-2{\rm i}z}J_0\left ( 2\sqrt{y z} \right )\Theta( z)\nonumber\\
&&+\Theta( z)\sqrt{y}\int_0^{ z} \rmd z^\prime \frac{\rme^{-{\rm
i}2z^\prime}}{\sqrt{z-z^\prime}}
I_1\left [ 2\sqrt{y(z-z^\prime)} \right ]
J_0\left (2\sqrt{yz^\prime} \right )\\
F_{1,1}(y,z)&=&\Theta( z)\int_0^{ z} \rmd z^\prime
\rme^{-{\rm i}2z^\prime}
I_0\left [ 2\sqrt{y(z-z^\prime)} \right ]
J_0\left ( 2\sqrt{y z^\prime} \right )\\
F_{2,0}(y,z)&=&\sqrt{\frac{z}{y}}I_1(2\sqrt{yz})\Theta(z)\nonumber\\
&&-\Theta(z)\int_0^z{\rm d}z^\prime \rme^{-2{\rm i}z^\prime}\sqrt{\frac{z-z^\prime}{z^\prime}}
I_1[2\sqrt{y(z-z^\prime)}]J_1[2\sqrt{yz^\prime}]
\\
F_{0,2}(y,z)&=&\rme^{-2{\rm i}z}\sqrt{\frac{z}{y}}J_1(2\sqrt{yz})\Theta(z)\nonumber\\&&+
\Theta(z)\int_0^{ z}{\rm d}z^\prime \rme^{-2{\rm i}z^\prime}\sqrt{\frac{z^\prime}{z-z^\prime}}
I_1[2\sqrt{y(z-z^\prime)}]J_1[2\sqrt{yz^\prime}]
\eea
\endnumparts
with $J_i$ and $I_i$ the $i$th Bessel function of the first kind and the $i$th modified Bessel function respectively \cite{BuNikZobLam}.
\end{appendix}

\end{document}